\documentclass[twocolumn,aps,prl,superscriptaddress,amsfonts]{revtex4-1}
\usepackage{mathrsfs,amsmath} 
\usepackage{graphicx}
\usepackage{bm}
\usepackage{array}
\usepackage{dcolumn}
\usepackage{hepparticles}
\usepackage{heppennames}
\usepackage{hepnicenames}
\usepackage{epstopdf}

\usepackage[usenames,dvipsnames]{xcolor}
\usepackage{soul}

\begin{document}

\def\e{\epsilon}
\def\d{\downarrow}
\def\u{\uparrow}
\def\e{\mathcal{E}}
\def\ba{\begin{eqnarray}}
\def\ea{\end{eqnarray}}
\def\beq{\begin{equation}}
\def\eeq{\end{equation}}
\title{High Resolution Adaptive Imaging of a Single Atom  }

\author{J. D. Wong-Campos}
\affiliation{Joint Quantum Institute, Joint Center for Quantum Information and Computer Science, and Department of Physics, University of Maryland, College Park, MD 20742}

\author{K. G. Johnson}
\affiliation{Joint Quantum Institute, Joint Center for Quantum Information and Computer Science, and Department of Physics, University of Maryland, College Park, MD 20742}

\author{B. Neyenhuis}
\affiliation{Joint Quantum Institute, Joint Center for Quantum Information and Computer Science, and Department of Physics, University of Maryland, College Park, MD 20742}

\author{ J. Mizrahi}
\affiliation{Joint Quantum Institute, Joint Center for Quantum Information and Computer Science, and Department of Physics, University of Maryland, College Park, MD 20742}

\author{ C. Monroe}
\affiliation{Joint Quantum Institute, Joint Center for Quantum Information and Computer Science, and Department of Physics, University of Maryland, College Park, MD 20742}

\date{\today}

\begin{abstract}
We report the optical imaging of a single atom with nanometer resolution using an adaptive optical alignment technique that is applicable to general optical microscopy.  By decomposing the image of a single laser-cooled atom, we identify and correct optical aberrations in the system and realize an atomic position sensitivity of $\approx$ 0.5 nm/$\sqrt{\text{Hz}}$ with a minimum uncertainty of 1.7 nm, allowing the direct imaging of atomic motion.  This is the highest position sensitivity ever measured for an isolated atom, and opens up the possibility of performing out-of-focus 3D particle tracking, imaging of atoms in 3D optical lattices or sensing forces at the yoctonewton (10$^{-24}$ N) scale.
\end{abstract}
\maketitle

The optical imaging of isolated emitters, such as individual molecules \cite{moerner_nobel,betzig_nobel}, optically active defects in solids \cite{nv_hell}, fluorescent dyes in a solution \cite{betzig}, or trapped atoms \cite{Greiner,BlattWineland08}, relies on efficient light collection and excellent image quality \cite{hell_nobel}.  Such high resolution imaging underlies many methods in quantum control and quantum information science \cite{Greiner,BlattWineland08}, such as quantum networks \cite{monroe0} fundamental atom-light interactions \cite{innsbruck}, and sensing small scale forces \cite{yocto}.  Individual atoms in particular have been resolved and imaged for many such applications \cite{grangier, jungsang,lucas, kielpinski1,boris}, with performance that depends critically on minimizing misalignments and optical aberrations from intervening optical surfaces such as a vacuum window.


In this article we develop a general method for suppressing aberrations by characterizing and adapting the imaging system, and report the highest performance optical imaging of an isolated atom to date. We image a single $^{174}$Yb$^{+}$ atomic ion with a position sensitivity of $\approx$ 0.5 nm$/\sqrt{\text{Hz}}$ for averaging times less than 0.1 s, observe a minimum uncertainty of 1.7(3) nm, and obtain direct measurements of the nanoscale dynamics of atomic motion. Complete knowledge on the wavefront distortions is obtained through the Zernike expansion of the point spread function and we adapt this information to correct aberrations and misalignments. The generality of the described work paves the way for adaptive optimal imaging in many other quantum optical systems as well as other contexts, such as biological microscopy or astronomy.

\section{Experimental Apparatus}

\begin{figure*}[ht]
 \includegraphics{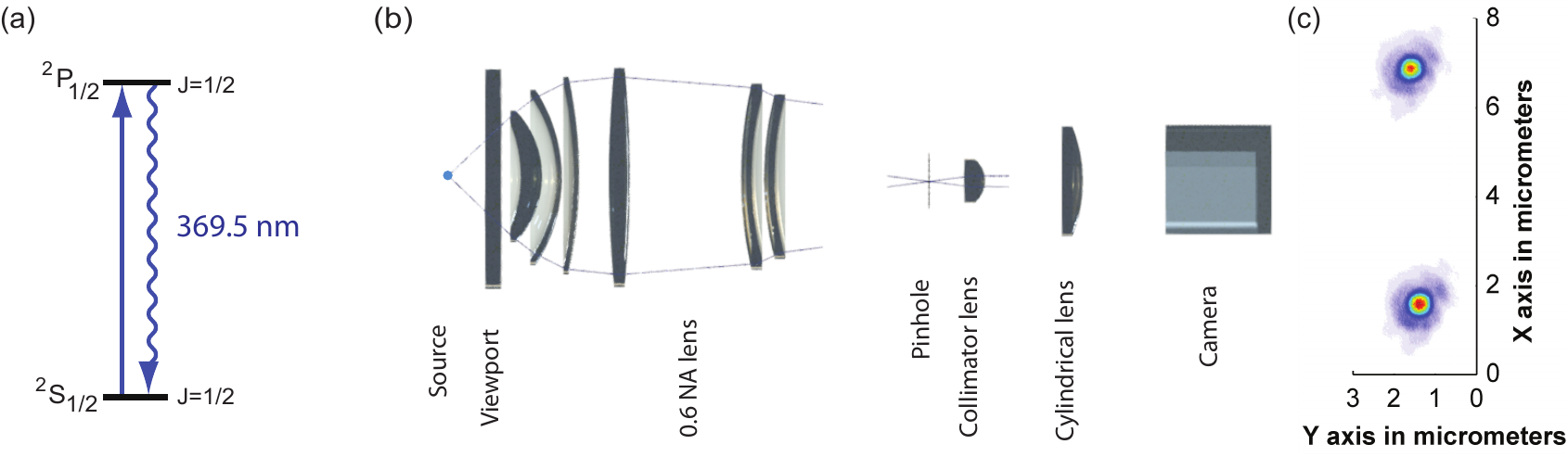} 
\caption{\textbf{Schematic of the imaging system}. (a) Atomic energy diagram of $^{174}$Yb$^{+}$. The atom is excited with laser radiation at 369.5 nm driving the  $^2S_{1/2} \rightarrow ^2P_{1/2}$ cycling transition and the resulting fluorescence is collected by the imaging system. (b) Transverse cut of the optical setup depicting the source, vacuum window, 0.6 NA objective lense, pinhole, short focal length lens, cylindrical lens and camera. (c) Image of two atomic ions separated by $\sim$ 5 $\mu$m.}
\label{fig:imgsyss}
\end{figure*}

The atomic imaging system is shown in Figure \ref{fig:imgsyss} (see also Supplementary section I). We confine a single $^{174}$Yb$^{+}$ ion in vacuum using a linear Paul trap \cite{leib,BlattWineland08} with 3D harmonic oscillation frequencies $(\omega_x, \omega_y, \omega_z)/2\pi = (1, 1.2, 0.8)$ MHz.  Laser light at a wavelength of $\lambda = 369.5$ nm is incident on the ion and resonantly excites the  $^2S_{1/2} \rightarrow ^2P_{1/2}$ cycling transition (radiative linewidth $\gamma/2\pi = 20$ MHz) as shown in fig. 1a.  The ion is laser-cooled and localized in each of the three dimensions of position to $\Delta x = \sqrt{(2\bar{n}+1)}x_0$, where $x_0= \sqrt{\hbar/2m\omega_x}\approx$ 5 nm is the zero-point spread, $\bar{n}$ is the mean thermal vibrational occupation number along each of the dimensions of motion, and $m$ is the atomic mass \cite{leib}.  For Doppler laser cooling with the cooling laser at an oblique angle to all directions of motion, $\bar{n} \approx \gamma/2\omega_x \sim 10$, thus $\Delta x \sim 20$ nm $\ll \lambda$ and the trapped ion acts as an excellent approximation to a point source.

The isotropic fluorescence from the atom at $\lambda = 369.5$ nm is transmitted through a vacuum viewport and collected by an objective lens of numerical aperture $\text{NA}=0.6$ with 10x magnification \cite{jungsang} (Figure \ref{fig:imgsyss}b).  The intermediate image passes through a pinhole that spatially filters light from background sources.  Additional magnification is provided by a second stage lens that forms an image at the face of an electron-multiplying-charge-coupled-device (EMCCD) camera (Figure \ref{fig:imgsyss}c).
The objective lens is mounted on a precision 5-axis alignment stage to compensate for comatic aberrations, and cylindrical optics are inserted after the magnifier lens to compensate for astigmatic aberrations.

\begin{figure*}[p]
\includegraphics{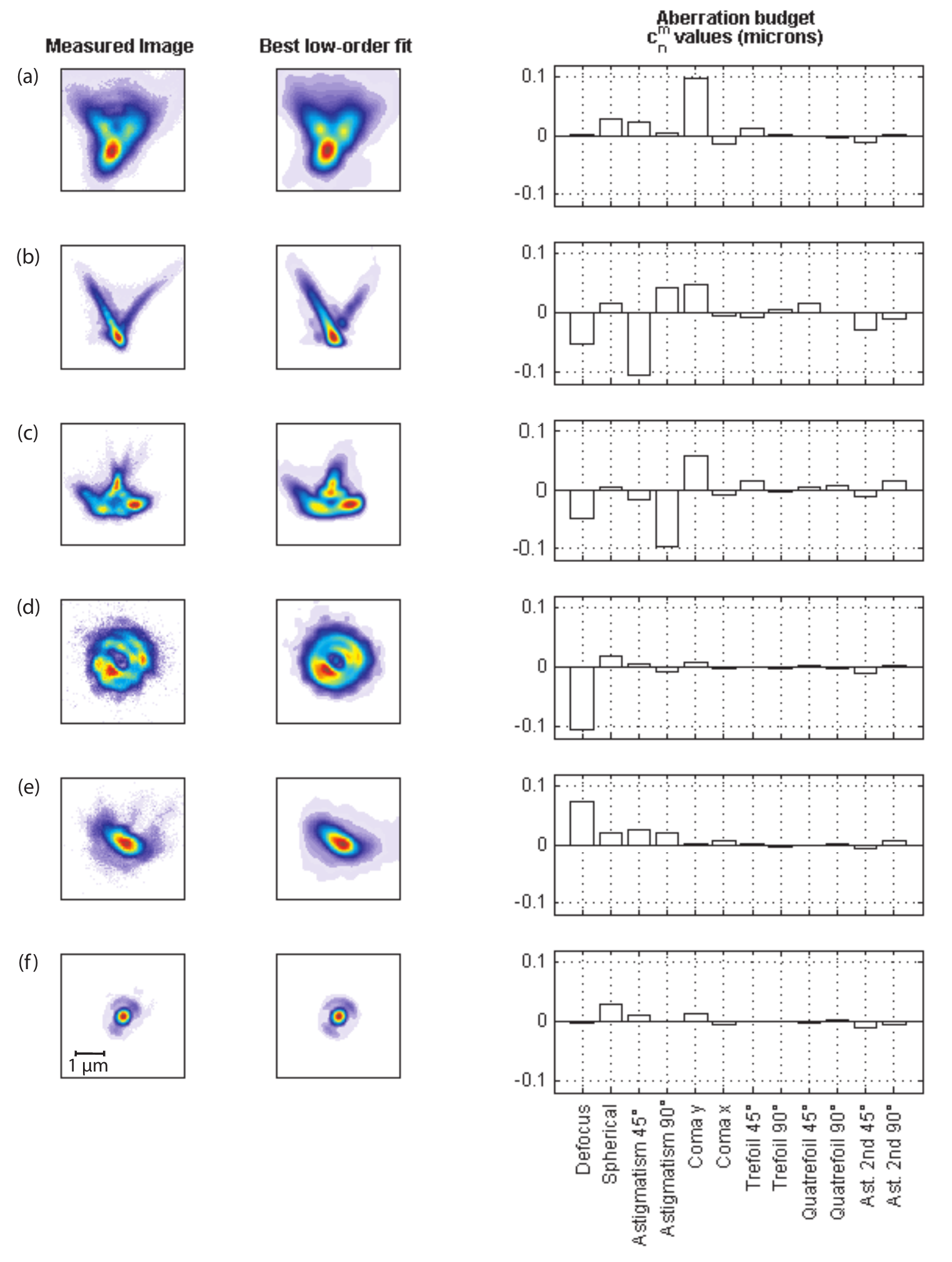} 
\caption{\textbf{Aberration retrieval results.} \textbf{(a,b,c)} Single shot images of the misaligned system. \textbf{(d,e,f)} depict the optimally aligned system at various distances from the focal plane, with (f) at the best focus. For \textbf{d} and \textbf{e} a high contribution from the defocus term is evident with low contributions of astigmatism and coma. Large contributions of coma and astigmatism (a-c) are corrected with a 5-axis stage and cylindrical lens (See supplemental material). The goodness of fit obtained for these examples approaches unity at coefficients of determination of 0.989, 0.965, 0.958, 0.957, 0.983 and 0.994 for images a,b,c,d,e and f respectively. These images are integrated for $\sim 0.5$ s.}
\label{fig:fit1}
\end{figure*}

\section{Aberration retrieval and suppression}
The measured spatial distribution of the image is the point spread function (PSF) \cite{goodman} which contains information about the ultimate resolution achievable in an imaging system and is the building block for more complex image formation through deconvolution techniques. The PSF can be decomposed into Zernike polynomials $Z^m_n(\rho,\theta)$ (See Methods)
in space\begin{equation}
\text{PSF}(\rho,\theta) =  \left|\mathscr{F}\left\{\exp \left(-i k\sum_{m,n} c^m_n Z^m_n(\rho,\theta)\right)\right\}\right|^2,\label{eq:psf}
\end{equation}
where $\mathscr{F}\{\}$ is the Fourier transform operator, $k = 2\pi/\lambda$ is the wavenumber \ and the $c^m_n$ coefficients are contributions of each Zernike component defined in the polar coordinates $\rho$ and $\theta$. The $c^m_n$ coefficients correspond to particular optical aberrations, so detailed characterization of the imaging system follows from the retrieval of the sign and magnitude of these coefficients.

Decomposing an image into Zernike polynomials relies on numerical algorithms \cite{pyramidal,retrieval} or semi-analytical calculations \cite{nijboer}. Here we obtain a full aberration characterization by using a least-squares fit to the measured data, using the $c^m_n$ coefficients and the exit pupil radius as fitting parameters. This represents a more general method for aberration retrieval since vector (polarization) effects for higher numerical apertures can be neglected \cite{novotny}. 

Figure \ref{fig:fit1} shows six single-shot images of a single $^{174}$Yb$^+$ ion. Figures
\ref{fig:fit1}a-c were taken during alignment and
Figs. \ref{fig:fit1}d-f were taken at different distances from the focal plane of the optimally aligned system. The images were integrated for $\sim 0.5$ s and fitted according to Eq. 1 to a linear superposition of the first twelve Zernike polynomial basis functions.  The overall fitting function is then smoothed by convolving with a Gaussian function that best fits the data and accounts for spatial drifts over long exposures. We find that the optimal image (Fig. 2f) has a characteristic radius of $\rho_0 = 363(18)$ nm, consistent with the diffraction-limited Airy radius of $\rho_0$ = 0.61$\lambda$/NA = 375.1 nm given the system numerical aperture.


Based on the one-to-one mapping of the Zernike polynomials to optical aberrations, we plot an aberration budget which shows the leading order aberration contributions to each of the images. For example, the contribution of the dominating negative (positive) defocus term of Fig. \ref{fig:fit1}d (Fig. \ref{fig:fit1}e) shows that we can map axial displacement on a transverse image distribution, with the position of best focus shown in Fig. \ref{fig:fit1}f. Moreover, a contribution of the comatic aberration indicates angular tilt errors and non-zero values of astigmatism indicate anisotropic foci in the system, seen in Figs. \ref{fig:fit1}a-c. 

This scheme provides a full quantitative basis for analyzing systems that rely on the aberrations introduced by the particle motion with respect to the objective to extract information on their dynamics. Examples of these experiments involve 3d off-focus tracking \cite{tracking1} and imaging of atoms arranged in 3d lattices \cite{weiss}. Although we describe an atomic emitter, this method can also be applied to the imaging of microbiological test samples (see supplementary material section III for an example).

\section{Position sensitivity}
\begin{figure*}[ht]
\includegraphics{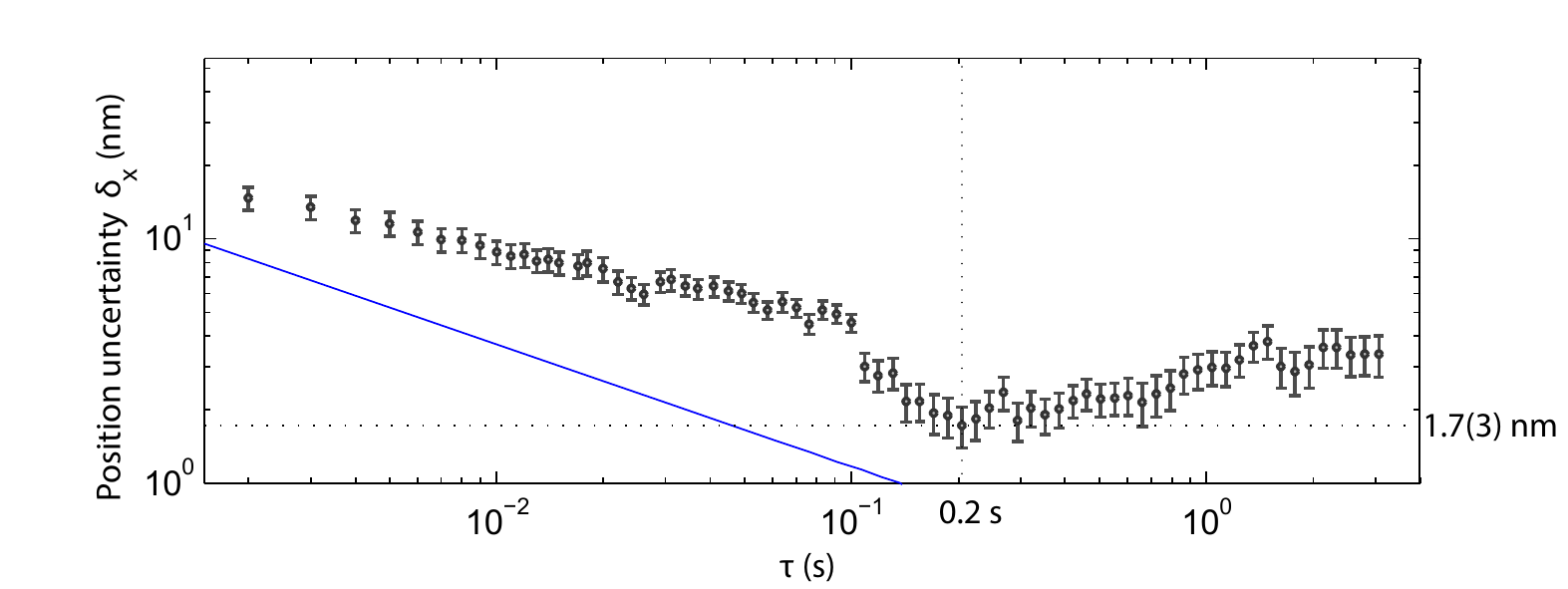} 
\caption{\textbf{Measured position uncertainty $\delta_x$ of the trapped ion centroid position versus image integration time $\tau$.} The blue line shows the expected uncertainty limited by photon counting shot noise in the imaging system. A sensitivity of  $\sim 0.5 $nm$/\sqrt{\text{Hz}}$ is measured for $\tau < 0.1$ s, which is $\sim 3$ times higher than shot noise, presumably from camera noise. The ultimate position sensitivity is found to be 1.7(3) nm at $\tau$ = 0.2 s.  These measurements include small corrections for dead time bias, as described in Methods.} \label{fig:allen}
\end{figure*}

The precision of measuring atomic position is dependent on the imaging system light collection and quality. As a result of the optical aberration characterization, even if it is not possible to directly correct the aberrations in the imaging system by alignment, it is feasible to post process and actively feed-forward the aberrated image and obtain a diffraction-limited performance through a digital filter with the phase information of the Zernike expansion. In this experiment we correct the aberrations by direct alignment (See supplementary information section I).

We measure the sensitivity on the position by taking $N$ images at 1 ms exposure time, binning them over total time duration intervals $\tau$ and calculating the Allan variance of the central position \cite{nist}.
\begin{equation}
\sigma^2 (\tau) = \frac{1}{2 (M-1)} \sum^{M-1}_{n=1} (y_{n + 1}-y_{n})^2,
\end{equation}
where M is the number of samples per bin and $y_{n}$ is the centroid of the ion image integrated over time $\tau$. Each image was integrated along one direction and fit to a one dimensional Gaussian linear count density function. The same procedure taken at different times $\tau$ leads to a curve of position uncertainty $\delta_x$ vs integration time as shown in Fig. \ref{fig:allen}. The data is corrected for a dead time of 5 ms between each 1 ms frame, allowing for state preparation and laser cooling (See methods and \cite{nist,barnes}).


The shot-noise-limited position sensitivity is given by $\sqrt{2\rho_0^2/R_0\tau}$, where $R_0 = \eta_D F \gamma/2$ is the maximum (saturated) measured fluorescence count rate from the atom, $F \approx 10\%$ is the solid angle fraction of fluorescence collected, and $\eta_D \approx 25\%$ is the quantum efficiency of the camera.  The observed sensitivity of $ \sim 0.5 $nm$/\sqrt{\text{Hz}}$ at small integration  times is somewhat higher than the expected level of shot noise (shown as the blue line in Fig. 3), and is consistent with observed super-Poissonian noise on the camera. Finite pixel size and background counts have negligible impact on the measured position sensitivity. We measure a minimum uncertainty of $\delta_x \approx 1.7(3)$ nm at an integration time of $\tau = 0.2$ s. For longer integration times, drifts in the relative position between the optical objective and the trapped ion degrade the position uncertainty as shown in Fig. \ref{fig:allen}, and with simple mechanical improvements in the imaging setup the resolution can likely be pushed well below 1nm.

Given this uncertainty in the position of the harmonically-bound ion, the sensitivity to detecting external forces is $\delta F = m\omega_x^2\delta_x$.  For a single $^{174}$Yb$^{+}$ ion with $\omega_x/2\pi = 10$ kHz, this would correspond to a force sensitivity in the yoctonewton ($10^{-24}$ N) scale, or an electric field at the $\mu$V/cm scale. Unlike earlier work 
\cite{yocto}, this imaging force sensor applies to single ions and does not require resolution of optical sidebands. 

\section*{Sensing of rf induced micromotion position}
\begin{figure*}[ht]
 \includegraphics{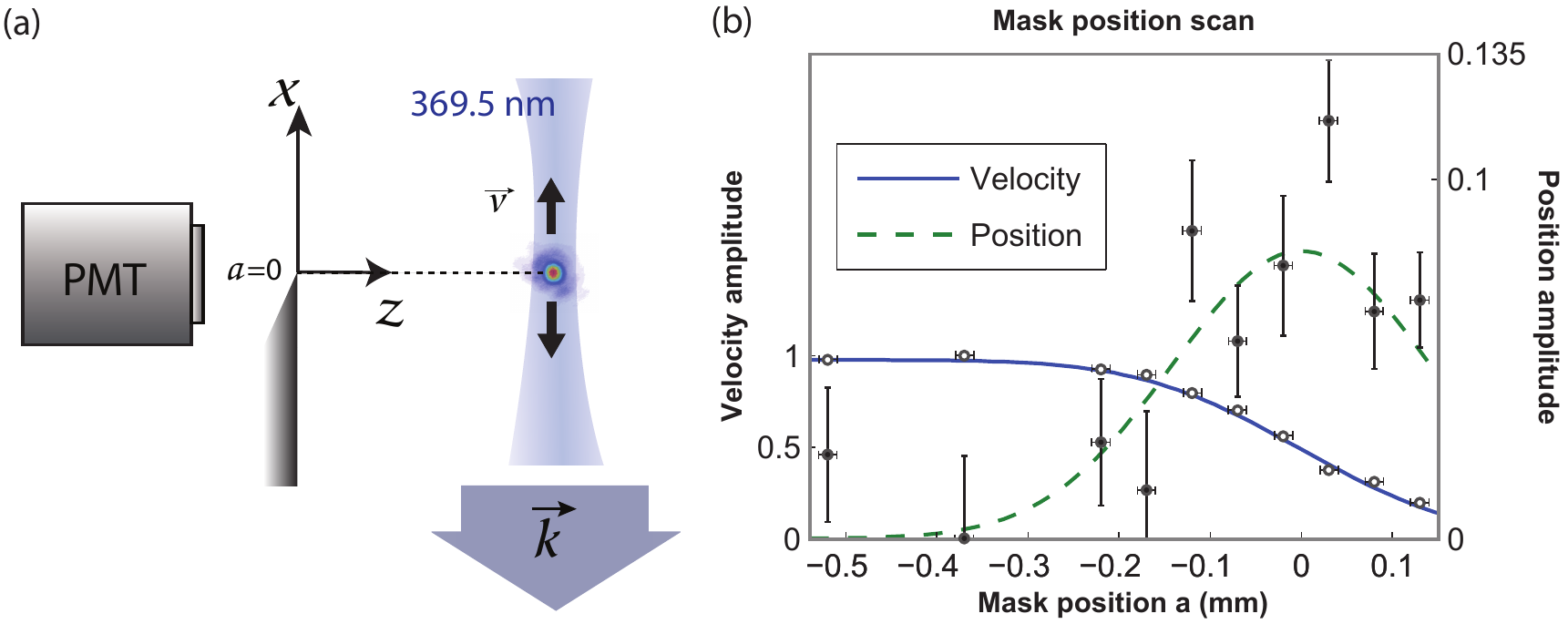} 
\caption{\textbf{Micromotion position measurement.} \textbf{(a)} The ion's velocity $\vec{v}$ (solid black arrows) is colinear with the direction $\vec{k}$ of the detection light, taken to be the $x$-axis. The fluorescence is modulated from the micromotion of the ion along $x$ by the first order Doppler effect  as well as the obscuraton by a mask with variable position $a$ along the $x$-axis. \textbf{(b)} Contributions of the velocity (solid blue line, left $y$ axis) and position (dashed green line, right $y$ axis) of a single atom when a mask is scanned along one transversal direction $x$. The values are normalized with the signal amplitude at $a = -\infty$.}
\label{fig:PMT}
\end{figure*}

Confinement of atomic ions in a Paul trap is achieved through oscillating rf electric field gradients that create a harmonic ponderomotive potential \cite{Dehmelt}.  In the presence of a static uniform electric field $E$, the ion acquires a ``micromotion" modulation in position 
$x(t) = X_{\mu}\sin\Omega t$ to first order in the pseudopotential approximation \cite{Dehmelt, leib}, where $\Omega$ is the drive frequency of the rf trapping field and $X_{\mu} = \sqrt{2}eE/(m\Omega\omega)$ is the micromotion amplitude.

The conventional approach for sensing micromotion is based on the first order Doppler modulation in the scattering of light from a laser beam of wavenumber $k$ propagating along the micromotion velocity \cite{micromot} (See fig \ref{fig:PMT}a). The correlation between the photon arrival times and the micromotion velocity is measured with a time-to-digital converter. With the excitation laser red-detuned from resonance of order $\gamma$ and for small levels of micromotion $kX_{\mu} \ll 1$, the measured fluorescence signal takes the form \cite{PTB},
\begin{equation}
R(t) = \alpha R_0 + \beta R_0 \left(\frac{k X_{\mu}\Omega}{\gamma}\right)\cos\Omega t ,
\end{equation}
where $\alpha,\beta \le 1$ are dimensionless constants that depend on the precise detuning and intensity of the excitation laser \cite{PTB}.

In order to also sense a direct position sensitivity to motion, we spatially mask the ion image with a sharp edge aperture, normal to the $(x)$ direction of motion. The mask position can be adjusted from, effectively, $a=-\infty$ (completely exposed) to $a=+\infty$ (completely masked) with $a=0$ covering exactly half of the image.  The total fluorescence behind the mask is then the integrated fluorescence behind the exposed area, 
\begin{multline}
R(a,t) = \alpha F(a) R_0 +\\ \beta F(a) R_0 \left(\frac{k X_{\mu}\Omega}{\gamma}\right)\cos\Omega t +\\ \alpha R_0 \frac{X_{\mu}}{\sigma\sqrt{\pi}} e^{-a^2/2\sigma^2} \sin\Omega t  , \label{eq:four}
\end{multline}
where we assume a Gaussian image distribution in space with root-mean-square radius $\sigma = 0.36\rho_0$ and the scale of the mask position $a$ is referred to the object.  The cumulative distribution function is $F(x)=[1-$erf$(\frac{x}{\sigma \sqrt{2}})]/2$.

We extract the two quadratures of the modulated fluorescence from Eq. \ref{eq:four} by performing sine and cosine transforms of the data.  The phases of the modulated signal are calibrated by opening the aperture $(a=-\infty)$ and taking the modulation as proportional to $\cos\Omega t$.

   Figure \ref{fig:PMT}b shows the position ($\sin\Omega t$) and velocity ($\cos\Omega t$) quadrature amplitudes (normalized to the amplitude at $a=-\infty$) as the mask position is scanned. 
   Based on the observed velocity-induced modulation in the count rate with full exposure ($a = -\infty$), we infer a micromotion amplitude of $X_{\mu}\sim$ 20 nm. As the mask is scanned along $x$, a position-dependent modulation in the fluorescence rate arises, reaching a maximum level at $a=0$. 
The absolute level of this position-dependent modulation is observed to be 15 times smaller than expected from Eq. \ref{eq:four}. This may be due  to slow drifts in the relative position of the ion with respect to the mask: a fluctuation of just 30 nm over the 300 s integration time required to obtain sufficient signal/noise ratio in the measurement would explain the observed reduction in the modulation.  

\section{Acknowledgements}
This work is supported by the U.S. Army Research Office (ARO) with funds from the IARPA MQCO Program and the ARO Atomic and Molecular Physics Program, the AFOSR MURI on Quantum Measurement and Verification,  the DARPA Quiness Program, the Army Research Laboratory Center for Distributed Quantum Information, the NSF Physics Frontier Center at JQI, and the NSF Physics at the Information Frontier program. The authors also acknowledge the support of the Imaging Core at the University of Maryland.

\section*{Methods}

\noindent \textbf{Aberration characterization}\\
Although optical aberrations can be described in terms of a Taylor expansion of the object height and pupil coordinates, Zernike polynomials $Z^m_n(\rho,\theta)$ are better suited since they form an orthogonal basis set of functions on a unit disk. Zernike polynomials are expressed in polar coordinates $\rho$ and $\theta$ as \cite{WYANT}

\begin{equation*} \label{eq:zernike}
Z^m_n(\rho,\theta)=\left\{
  \begin{array}{l l}
   N^m_n R^m_n(\rho)\cos(m\theta) & \quad \text{for m $\geq$ 0}\\
   N^m_n R^m_n(\rho)\sin(m\theta) & \quad \text{for m $<$ 0},
  \end{array} \right.
\end{equation*} 
\begin{equation*}
N^m_n = \sqrt{\frac{2(n+1)}{1+\delta_{m0}}},
\end{equation*}
\begin{multline}
R^{|m|}_n(\rho) =  \sum\limits_{s=0}^{(n-|m|)/2}\frac{(-1)^s}{s! [(n+|m|)/2 - s ]!} \\ \times \frac{(n-s)!}{[(n-|m|)/2 - s]!}\left(\frac{\rho}{\rho_p}\right)^{n-2s},
\end{multline}

\noindent where $n$ is an integer number and $m$ can only take values $n, n-2, n-4, ..., -n $ for each $n$. The radial coordinate is scaled to the exit pupil radius $\rho_p$ (the radius of the image of the input aperture at the camera). Importantly, each term of this polynomial expansion has a one-to-one relation with a specific kind of aberration. Given the Zernike expansion of a wavefront, we can calculate its deviation from a perfect  wavefront using the $c^m_n$ coefficients of eq. (\ref{eq:psf}). \\

%

\noindent 
\textbf{Dead time corrections}\\
Dead times were corrected using the Allan B-functions \cite{barnes} defined in the suplementary section
\begin{equation}
\sigma^2(\tau) = \frac{\sigma^2(2,\mathcal{M}T_0,\mathcal{M}\tau_0)}{B_3(\mu)B_2(\mu)} \label{eq:correction}
\end{equation}
where $\mu$ is the noise model coefficient that range between $-1<\mu<1$, $\mathcal{M}$ is the binning parameter, $T_0$ is the time between data acquisitions and $\tau_0$ is the sampling time. Dead times are then defined as $t_{dead}=T_0-\tau_0$ for single acquisition times. The integration time for the Allan variance is $\tau = \mathcal{M}\tau_0$. The noise model $\mu$ upon which the B-functions depend at each $\tau$ were found solving 
\begin{equation}
\frac{B_1(\mu)}{B_3(\mu)}=\frac{\sigma^2(N,T,\tau)}{\sigma^2(2,\mathcal{M}T_0,\mathcal{M}\tau_0)}
\end{equation}
for $\mu$ with $\sigma^2(N,T,\tau)$ defined as the standard variance.


\onecolumngrid
\newpage 
\section{Supplemental material}
\subsection{Section I - Details of the experimental set-up}
The atomic ion is positioned 11.6 mm  from a 4 mm vacuum window. The first assembly of six lenses fabricated by Photon Gear, Inc. allows collection from a large numerical aperture with near diffraction-limited performance \cite{jungsang}.  After this lens, we place a 128 $\mu$m pinhole to spatially filter the scattered light from the ion trap followed by a short focal length lens. The final measured magnification of the system is 471(3), which was measured by moving the stage holding the lens assembly and observing the translation of the image. As the detector, we used an iXon Ultra 897 electron multiplying camera (EMCCD) with pixel size of 16 $\mu$m.
Because angular alignment of the lens assembly is crucial, we corrected the tilt with a 5-axis alignment stage with angular resolution of 100$\mu$rad. The depth of focus was measured to be on the order of 0.5$\mu$m. The presence of astigmatism, which may stem from clamping of the imaging system or cylindrical warping of the vacuum glass, was corrected by placing a slow cylindrical lens after the last lens. The performance of this design was simulated in ZEMAX with an aberration-free spot size of $374.6$nm. 

\begin{figure*}[ht]
\begin{center}
\centering
 \includegraphics{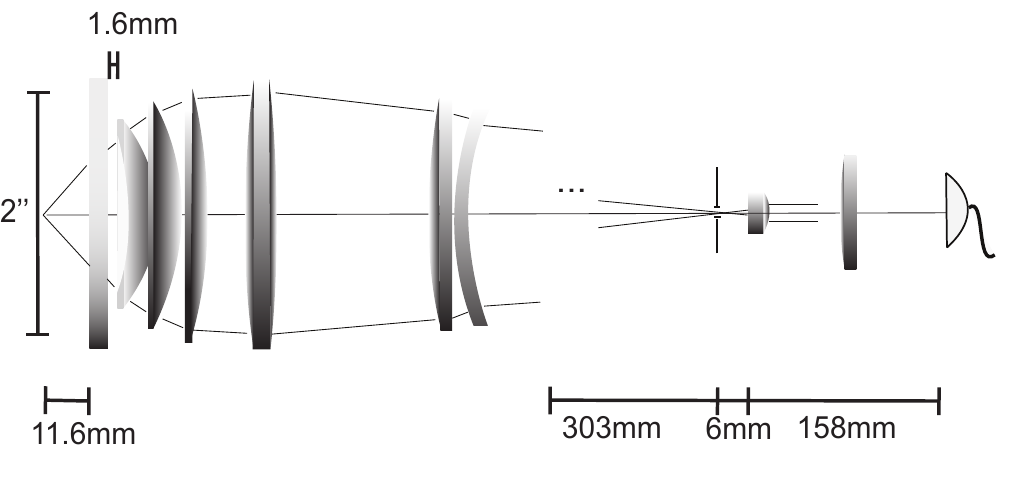} 
\end{center}
\caption{Imaging system. A 4 mm vacuum viewport isolates the atomic ion from the environment. Corrections to the aberrations were performed through the 5-axis translational stage and cylindrical lens.}
\end{figure*}

\subsection{Section II - Aberration characterization on a state of the art microscope for microbiology research. }

Widefield fluorescence images of microspherical test specimens of 0.5 $\mu m$ radius are taken with the 60x lens array of a DeltaVision Elite (Applied Precision) widefield microscope with $ \text{NA}= 1.49$ . The images are expanded in the Zernike basis as described in the paper. We show three images at three different focal planes, the coefficients of determination obtained were 0.77 0.82 and 0.91.\\

\begin{figure}[ht]
\begin{center}
\centering
 \includegraphics{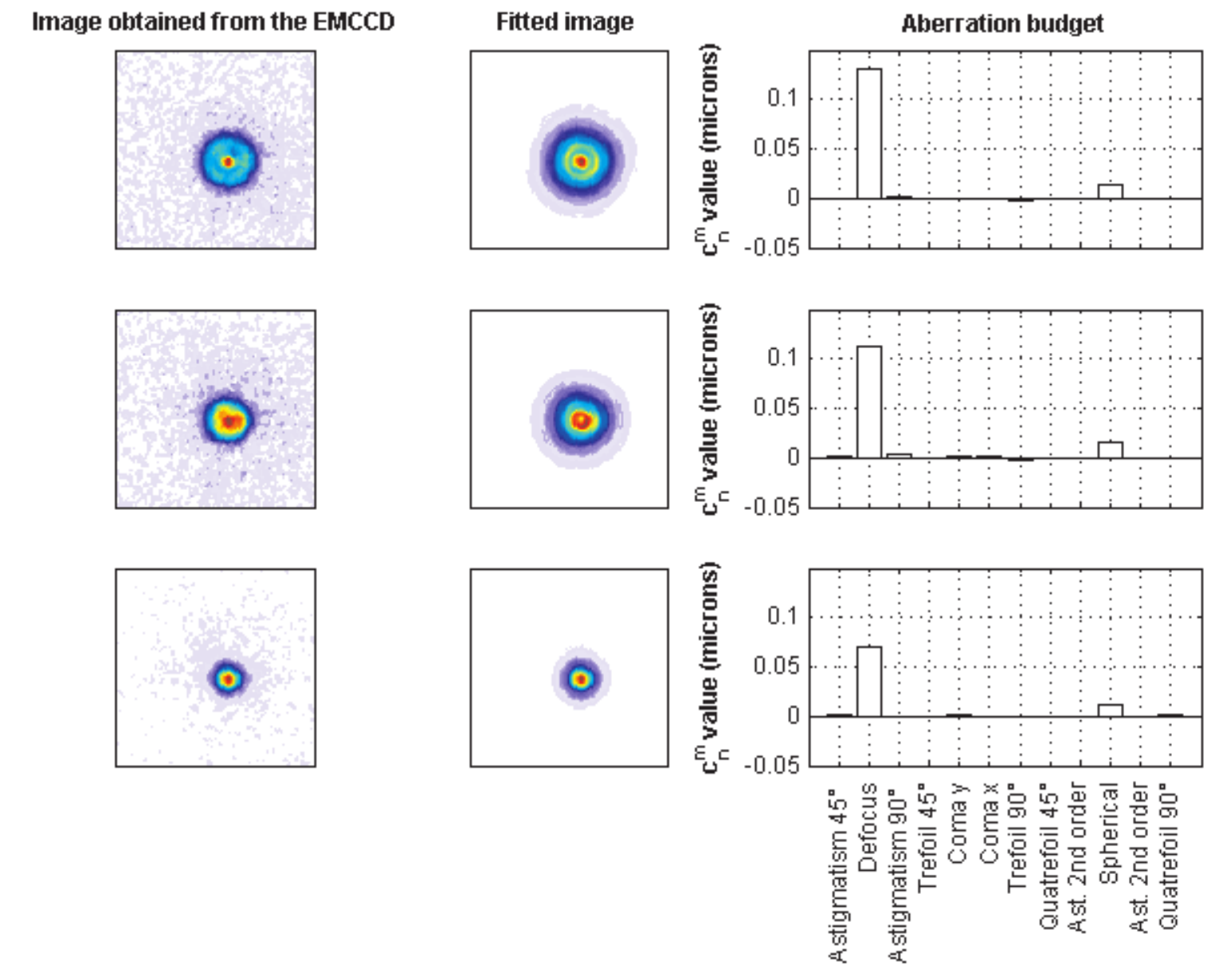} 
\end{center}
\caption{Images of fluorescent microspherical specimen of radius 0.5 $\mu$m taken with a DeltaVision Elite widefield microscope. Defocus and spherical aberrations are recognized as the largest contribution in the apparatus. The errors on the fit are caused by the spurious noise of the data.}
\end{figure}

\newpage

\subsection{Section III - Allan deviation dead time analysis}
\noindent Dead times in the experiment were corrected introducing the Allan B-functions \cite{barnes}:\\
We first define the Bias functions with the help of the function

\begin{equation}
F(A)=2A^{\mu+2}-(A+1)^{\mu+2} - |(A-1)|^{\mu+2}
\end{equation}

\begin{center}
  \begin{tabular}{| l | r |}
    \hline
    Noise &  $\mu$ \\ \hline
    White &  -1 \\ \hline
    Flicker & 0 \\\hline
    Random walk & 1 \\
    \hline
  \end{tabular}
\end{center}

\noindent \textbf{The $B_1$ Bias function }\\

\begin{equation}
B_1(N,r,\mu)=\frac{\sigma^2(N,T,\tau)}{\sigma^2(2,T,\tau)} = \frac{1+\sum_{n=1}^{N-1}\frac{N-n}{N(N-1)}F(nr)}{1+(1/2)F(r)}
\end{equation}
This coefficient relate the standard variance $\sigma^2(N,T,\tau)$ with the Allan variance including dead times at the end of the measurement (Without binning).\\

\noindent \textbf{The $B_2$ Bias function }\\

\begin{equation}
B_2(r,\mu)=\frac{\sigma^2(2,T,\tau)}{\sigma^2(2,\tau,\tau)}=\frac{1+(1/2)F(r)}{2(1-2^{\mu})}
\end{equation}
This coefficient relate the Allan variance with dead time  $\sigma^2(2,T,\tau)$ with the Allan variance free of dead times $\sigma^2(2,\tau,\tau)$\\

\noindent \textbf{The $B_3$ Bias function for a two sample variance }

\begin{align}
B_3(2,\mathcal{M},r,\mu)&=\frac{\sigma^2(2,\mathcal{M}T_0,\mathcal{M}\tau_0)}{\sigma^2(2,T,\tau)}\\
&=\frac{2\mathcal{M}+F(\mathcal{M}r)\mathcal{M}-\sum_{n=1}^{\mathcal{\mathcal{M}}-1}(\mathcal{M}-n)[2F(n r)-F((\mathcal{M}+n)r)-F((\mathcal{M}-n)r)]}{\mathcal{M}^{\mu+2}[F(r)+2]}
\end{align}
This coefficient relate the Allan variance with periodic dead times $\sigma^2(2,\mathcal{M}T,\mathcal{M}\tau)$ where $\mathcal{M}$ is the binning parameter and the Allan variance with dead times accumulated at the end of the sampling is $\sigma^2(2,T,\tau)$.\\
In this experiment, we binned $M =$ 200 images of $\tau_0 =$ 1 ms with a dead time of 5 ms obtaining $T_0 =$ 6 ms. That is, we measured $\sigma^2(2,\mathcal{M}T_0,\mathcal{M}\tau_0)$. To obtain a dead time corrected Allan variance we need to:
\begin{equation}
\sigma^2(2,\tau,\tau) = \frac{\sigma^2(2,\mathcal{M}T_0,\mathcal{M}\tau_0)}{B_3B_2}
\end{equation}
Or the standard deviation\\
\begin{equation}
\sigma(2,\tau,\tau) = \frac{\sigma(2,\mathcal{M}T_0,\mathcal{M}\tau_0)}{\sqrt{B_3B_2}} \label{eq:correction}
\end{equation}

\section*{Noise detection}
To obtain a noise model we use the $B_1$ bias functions

\begin{equation}
B_1(N,r,\mu)=\frac{\sigma^2(N,T,\tau)}{\sigma^2(2,T,\tau)}
\end{equation}
For this we need the $\sigma^2(2,T,\tau)$ which can be obtained from the $B_3$ functions:
\begin{equation}
\frac{B_1(N,r,\mu)}{B_3(N,r,\mu)}=\frac{\sigma^2(N,T,\tau)}{\sigma^2(2,MT_0,M\tau_0)}
\end{equation}
To obtain the standard deviation we then square root this expression and solve this equation for $\mu$
\begin{equation}
\sqrt{\frac{B_1(N,r,\mu)}{B_3(N,r,\mu)}}=\frac{\sigma(N,T,\tau)}{\sigma(2,MT_0,M\tau_0)}
\end{equation}
We replace the obtained $\mu$ in equation (\ref{eq:correction}) to find the Allan deviation correction with dead times.

\end{document}